\documentclass[12pt,onecolumn]{article}
\usepackage[a4paper,margin=1in]{geometry}

\usepackage{cite}
\usepackage{amsmath,amssymb,amsfonts}
\usepackage{algorithmic}
\usepackage{graphicx}
\usepackage{textcomp}
\usepackage{xcolor}
\usepackage{xspace}
\usepackage{multirow}
\usepackage{lipsum}
\usepackage{stfloats}
\usepackage{comment}
\usepackage{subcaption}
\usepackage{url}

\newcommand{\sysname}{\textsc{DSDE}\xspace}
\def\BibTeX{{\rm B\kern-.05em{\sc i\kern-.025em b}\kern-.08em
    T\kern-.1667em\lower.7ex\hbox{E}\kern-.125emX}}

\title{\textbf{DSDE: Dynamic Speculative Decoding with KLD Stability for Real-World Serving}}

\author{\large Mingyu Yang$^{\dagger}$, Jae-Young Choi$^{\dagger}$, Kihyo Moon, Minsung Jang, Eunjoo Jeon$^{*}$\\[4pt] \footnotesize Cloud Research Team, Samsung SDS, Seoul, Republic of Korea}

\date{}

\begin{document}
\maketitle

\begingroup
\renewcommand\thefootnote{\fnsymbol{footnote}}
\footnotetext[2]{These authors contributed equally.}
\footnotetext[1]{Corresponding author: \texttt{ej85.jeon@samsung.com}}
\footnotetext[3]{%
This paper has been accepted for presentation at the 
\textit{11th Special Session on Intelligent Data Mining, 
IEEE International Conference on Big Data (BigData 2025)}.%
}
\footnotetext[4]{%
© 2025 IEEE. Personal use of this material is permitted. 
Permission from IEEE must be obtained for all other uses, including reprinting or republishing for advertising or promotional purposes, creating new collective works, for resale or redistribution to servers or lists, or reuse of any copyrighted component of this work in other works.%
}
\endgroup

\begin{abstract}
Speculative decoding accelerates large language model inference, but its reliance on a fixed speculation length is suboptimal in large-batch serving environments with diverse requests. This paper explores a new direction for dynamic adaptation by investigating a novel class of post-hoc, diagnostic signals. We propose Dynamic Speculative Decoding Engine (\sysname), a training-free framework built on two primary components: (1) a predictive signal based on the variance of the Kullback-Leibler (KLD) divergence, which diagnoses the generation’s regional stability, and (2) an adaptive speculation length cap to mitigate the straggler problem in per-sequence decoding. Experiments demonstrate the potential of using KLD-based stability signals for dynamic adaptation. An algorithm guided by these signals achieves end-to-end latency competitive with leading baselines and exhibits superior robustness across diverse workloads. This robustness is particularly valuable in challenging low-acceptance-rate regimes, where the proposed signal maintains its diagnostic utility. Collectively, these findings validate post-hoc signals as a valuable component for building more robust and intelligent LLM inference systems, and highlight a promising direction for future research on dynamic speculation length adaptation.
\end{abstract}


\section{Introduction}
\label{sec:Introduction}

Large Language Models (LLMs) have become central to modern NLP, but their practical deployment in latency-sensitive services remains constrained by slow auto-regressive decoding. Although GPUs provide high throughput, generation proceeds token-by-token, which limits hardware utilization and increases per-sequence latency. LLM inference is fundamentally memory-bound, as each decoding step repeatedly accesses the model weights and the KV cache. This leads to low arithmetic intensity and under-utilization of compute resources~\cite{shazeer2019fast}. To mitigate this, batching is widely adopted~\cite{narayanan2021megatron,dao2022flashattention}, which improves GPU utilization by shifting the bottleneck toward compute. However, batching alone does not reduce the inherent latency of any single sequence and can even introduce queuing delays in latency-sensitive environments~\cite{sheng2024slora}. This motivates complementary algorithmic strategies that directly reduce decoding latency. Speculative decoding has emerged as one of the most effective such strategies~\cite{chen2023specdecode}. A lightweight draft model proposes multiple candidate tokens, and the target model verifies them in parallel. This approach reduces the number of target model inferences and improves decoding efficiency. Higher acceptance rates yield fewer decoding steps and greater speedups. However, current implementations typically use a static speculation length (SL) across all sequences and decoding steps. This rigidity is problematic in heterogeneous large-batch serving: mixed tasks such as code generation and dialogue require different speculation strategies (Table~\ref{tab:strategy_comparison}). A single static SL forces the entire batch to follow the least efficient sequence, creating a scalability bottleneck. This limitation motivates a more flexible approach, in which speculation length is assigned per sequence within a batch.

\begin{table}[t]
\centering
\tiny 
\setlength{\tabcolsep}{4pt} 
\renewcommand{\arraystretch}{1.3} 
\caption{Performance of Static SL Strategies on Heterogeneous Tasks (HumanEval vs. ShareGPT),
showing latency (second) and block efficiency (BE) differences between $SL=2$ and $SL=8$.}
\resizebox{0.9\textwidth}{!}{%
\begin{tabular}{|p{1.4cm}|p{4.0cm}|p{1.4cm}|p{1.4cm}|}
\hline
\textbf{Task} & \textbf{Speculation Strategy} & \textbf{Latency} & \textbf{BE} \\
\hline
\multirow{2}{*}{Code} 
    & Static-Aggressive ($SL=8$) & 15.92 & 5.87 \\
\cline{2-4}
    & Static-Conservative ($SL=2$) & 21.56 & 2.67 \\
\hline
\multirow{2}{*}{Dialogue} 
    & Static-Aggressive ($SL=8$) & 19.27 & 4.81 \\
\cline{2-4}
    & Static-Conservative ($SL=2$) & 22.24 & 2.54 \\
\hline
\end{tabular}
}
\label{tab:strategy_comparison}
\end{table}

\begin{figure*}[t]
    \centering
    \includegraphics[width=1.0\textwidth]{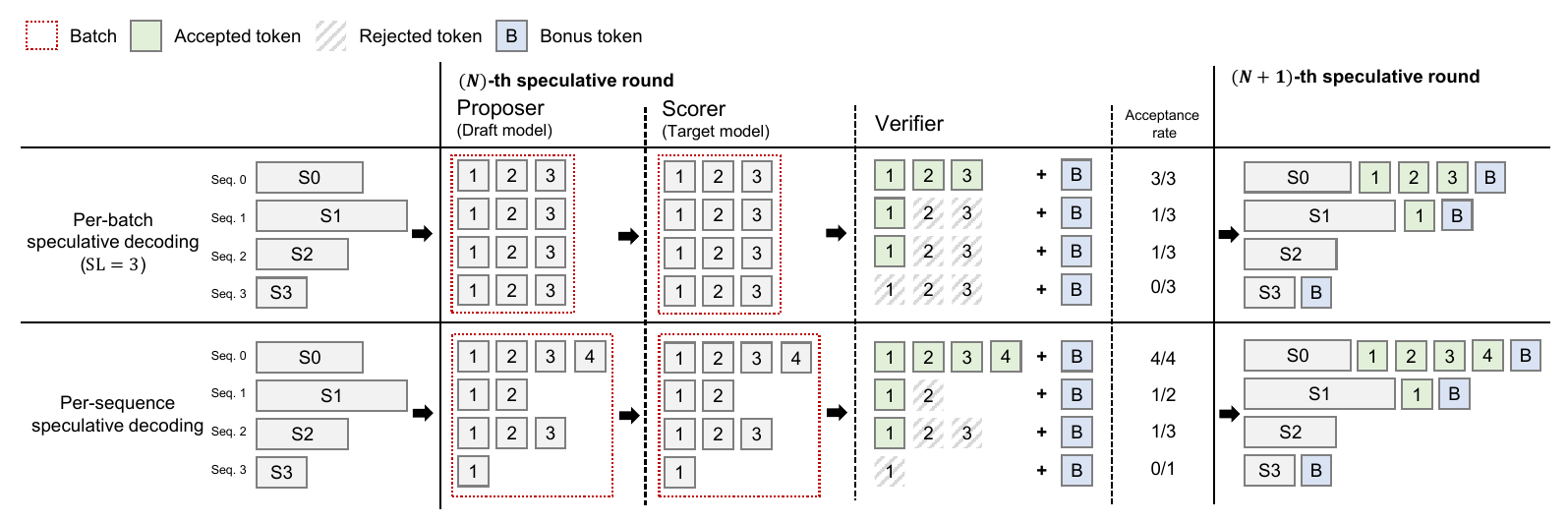}
    \caption{Two strategies for speculative decoding: per-batch decoding with a static speculation length (top, $SL=3$) and per-sequence decoding with adaptive speculation lengths (bottom).}
    \label{fig:sl_motivation}
\end{figure*}

Figure~\ref{fig:sl_motivation} illustrates the key difference between these two strategies. Assuming a batch size of four, the top panel depicts the per-batch approach where a static SL (e.g., $SL=[3, 3, 3, 3]$) is uniformly applied to all sequences, irrespective of their individual contexts. In contrast, the bottom panel shows the per-sequence approach. Here, the speculation length is chosen per sequence (e.g., $SL=[4, 2, 3, 1]$), tailored to their anticipated characteristics. This per-sequence assignment offers several advantages. First, it allows for adaptive control over the draft model's generation length; SL can be increased for sequences with a high expected acceptance rate to generate more tokens per verification step, or decreased to prevent wasteful computation on tokens likely to be rejected. Second, it can improve resource management within the batch: sequences that complete their generation early can be removed, allowing computational resources to be reallocated to the remaining active sequences.

However, even a per-sequence approach has limitations. The optimal speculation length for a single sequence is not constant; it changes over time depending on the context of the generation. Assigning a static SL to a sequence, even if tailored, fails to adapt to these fluctuations. Therefore, a truly efficient solution must be dynamic, adjusting the speculation length on-the-fly for each sequence at each decoding step.

\begin{figure}[t]
    \centering
    \begin{minipage}[t]{0.48\textwidth}
        \centering
        \includegraphics[width=\linewidth]{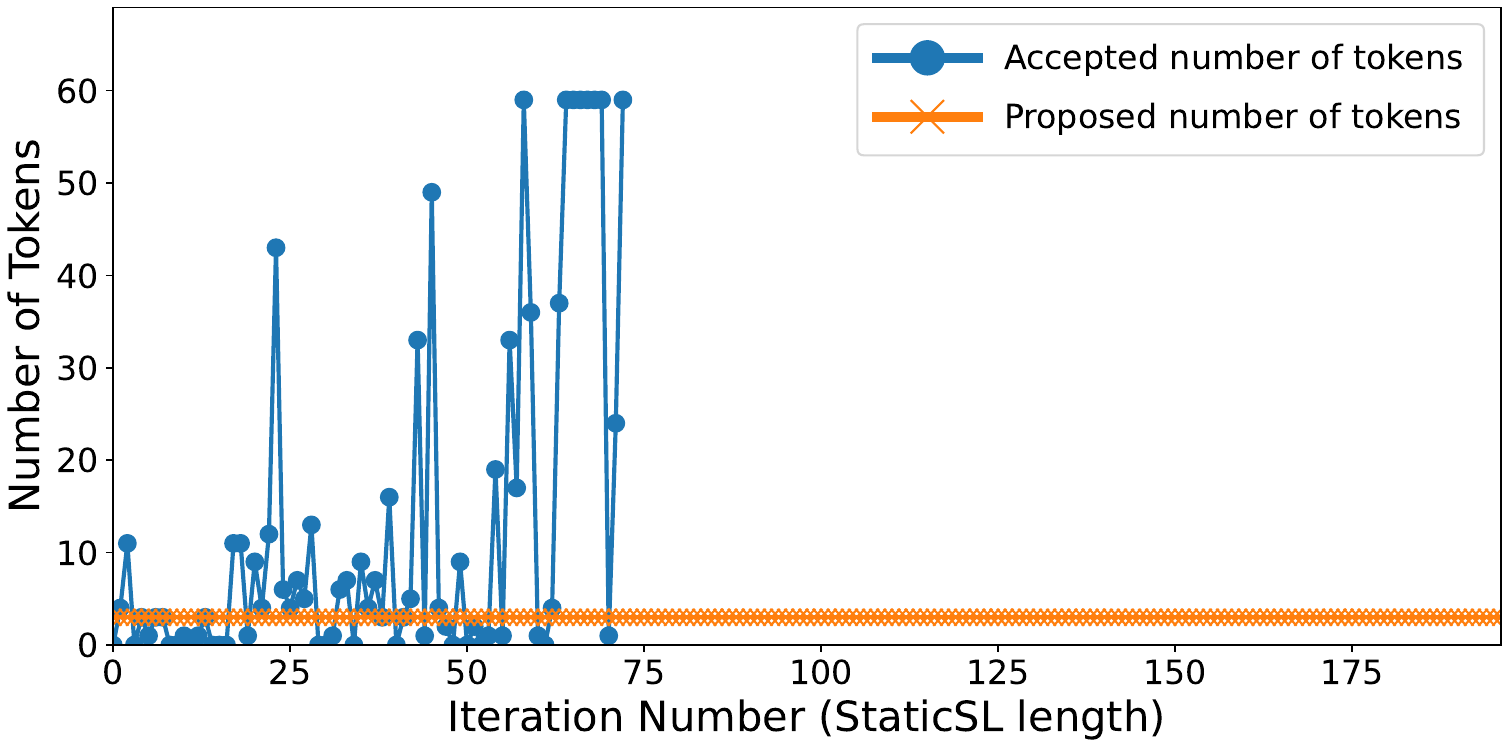}
        \vspace{-3pt}
        \caption{Iteration-level fluctuation of the optimal speculation length, highlighting the challenge for dynamic prediction.}
        \label{fig:sl_fluctuation}
    \end{minipage}
    \hfill
    \begin{minipage}[t]{0.48\textwidth}
        \centering
        \includegraphics[width=\linewidth]{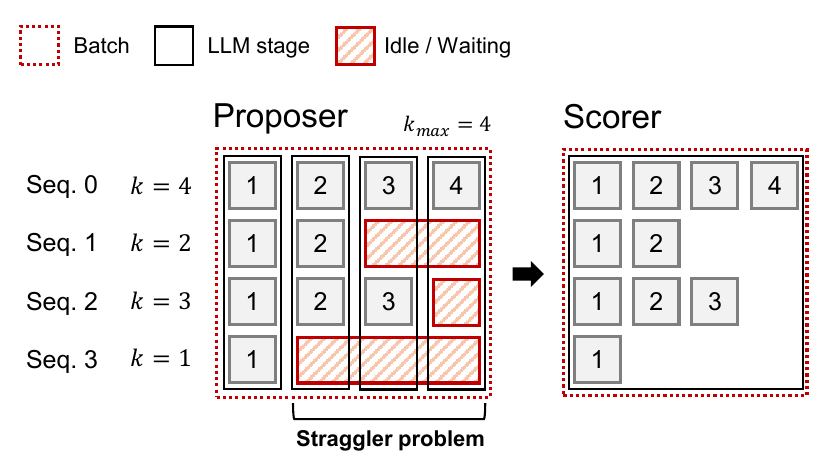}
        \vspace{-3pt}
        \caption{Straggler problem in a per-sequence approach.}
        \label{fig:straggler}
    \end{minipage}
    \vspace{-5pt}
\end{figure}

Designing an effective and practical dynamic per-sequence and per-iteration system presents two primary challenges:
\begin{itemize} 
\item \textbf{Training-free requirement.} The highly volatile nature of the optimal SL (Figure~\ref{fig:sl_fluctuation}) has led many existing strategies to rely on dedicated, trained models. While often performance, these models introduce significant overhead, requiring costly retraining for new target models or datasets. A practical solution should therefore be lightweight and avoid retraining.

\item \textbf{Straggler effect in large-batch serving.} When applying varying per-sequence lengths within a batch, a naive implementation forces faster sequences (those with a short SL) to wait idly for the longest prediction to complete. As illustrated in Figure~\ref{fig:straggler}, this synchronization bottleneck can diminish the benefits of per-sequence adaptation, especially as batch sizes increase. 
\end{itemize}

To address these, we propose a training-free method and a mechanism to resolve the straggler problem. The key insight is that simple forward-looking metrics like the draft model's entropy often correlate only weakly with token acceptance, limiting their predictive utility (for example, on the CNNDM dataset, the Pearson correlation between entropy and acceptance rate using a LLaMA-3.2 1B draft model was $r=-0.339$ with $p<0.001$).

Instead, we investigate \emph{post-hoc}, lagging signals obtained after the target model’s verification. Among these, the Kullback–Leibler divergence (KLD), which is a measure of model disagreement, is particularly informative. Rather than raw KLD, we hypothesize that its \emph{variance} across recent steps indicates \emph{regional stability}: whether the current generation phase is predictable or difficult. To mitigate the straggler problem, we introduce an adaptive speculative-length cap, $SL_{\text{cap}}$, a dynamic upper bound applied across the batch to balance aggressive per-sequence speculation with high throughput for the batch as a whole.

Therefore, we propose \sysname, the Dynamic Speculative Decoding Engine, which adjusts SL per sequence and per iteration. It operates online, does not require retraining, and integrates seamlessly into vLLM’s speculative decoding pipeline\cite{kwon2023efficient}. This enables efficient large-batch inference, improving latency, block efficiency, and goodput over strong static SL baselines. In summary, our contributions are as follows:
\begin{itemize} 
\item \textbf{Problem identification:} We characterize the inefficiency of static SL in large-batch decoding, motivating the need for dynamic adaptation.
\item \textbf{Training-free online adapter:} A lightweight algorithm that adjusts SL per sequence and iteration using KLD variance (optionally combined with entropy).
\item \textbf{vLLM-based integration:} We implement \sysname within vLLM’s speculative decoding architecture, supporting real-world deployment without model modification.
\item \textbf{Comprehensive evaluation:} We demonstrate consistent improvements in latency and block efficiency across models and workloads compared to static SL baselines. 
\end{itemize}

\section{Related Work}
\label{sec:related_work}

Research on speculative decoding can be understood from two complementary perspectives: 
the \textit{architectural paradigm} that defines how draft tokens are generated, 
and the algorithmic or system-level optimizations built on top of these paradigms. 
We first review the two dominant speculative decoding architectures.

\subsection{Architectural Paradigms of Speculative Decoding}

\textbf{Target-Internal Multi-Head Drafting.}  
One line of work equips the target model itself with additional decoding heads that directly predict multiple subsequent tokens in parallel. Medusa~\cite{medusa2024} exemplifies this design: it augments the base LLM with extra heads and a tree-based attention mechanism, constructing multiple candidate continuations that are verified simultaneously within each decoding step, thereby reducing decoding rounds. Eagle~\cite{eagle2024} follows a similar philosophy. DEL~\cite{del2025} further enhances this paradigm by dynamically selecting exit layers and speculation lengths within the target model at inference time, enabling context-aware draft–verify tradeoffs.

\textbf{Small Draft + Large Target Model.}  
Another widely adopted paradigm employs a lightweight draft model alongside a large target model. This approach was independently proposed by Leviathan et al.~\cite{leviathan2023sd} and Chen et al.~\cite{chen2023specdecode}, and has become the basis for many dynamic SL approaches—which we elaborate on in the subsequent sections.

\subsection{Algorithmic Enhancements for Dynamic SL}

\textbf{Training-based Approaches.} A prominent strategy involves training an auxiliary model to guide the speculation process. DISCO~\cite{mamou2024disco}, for instance, trains a classifier to predict a confidence score for early stopping. Similarly, SPRINTER~\cite{zhong2025sprinter} trains a lightweight verifier that performs approximate verification, invoking the target model only upon a predicted rejection, though this relaxes the exact-match constraint. While effective, these methods introduce the overhead of a dedicated training phase.

\textbf{Training-free Approaches.} As a more lightweight alternative, training-free heuristics use pre-defined formulas on signals from the draft model. AdaEDL~\cite{agrawal2024adaedl}, for example, uses the entropy of the draft model's logits as an early-stopping criterion.

\subsection{Stability as a Signal in Language Generation}
The concept of using signal stability to guide the generation process was explored in the foundational work on the Stable Entropy Hypothesis by Arora et al.~\cite{arora2023stable}. They postulate that high-quality text generation maintains its conditional entropy within a narrow 'stable entropy zone,' and their experiments empirically demonstrate that deviations from this zone correlate with generation failures like repetition or incoherence. Based on this, they developed Entropy-Aware Decoding (EAD), an algorithm that actively intervenes to improve output quality by keeping the generation within these stable bounds. 

Building on this line of entropy-based adaptation, SVIP~\cite{svip2025} derives a theoretical lower bound on acceptance probability by linking draft entropy with KL–TVD inequalities, yielding a practical entropy-thresholding rule for draft length control. 
In parallel, DySpec~\cite{dyspec2025} leverages the draft model’s probability mass as a proxy for acceptance rate, showing that tokens with higher draft probabilities are more likely to be accepted. While DySpec briefly mentions KL divergence to justify the closeness between draft and target distributions, KLD is not directly used as a runtime signal.

Our work is conceptually aligned with leveraging signal stability, but targets a more direct signal for speculative decoding. While Arora et al.~\cite{arora2023stable} analyze the stability of a single model’s internal state via its entropy, and SVIP connects entropy to acceptance while DySpec bridges draft probability with acceptance, our approach measures the stability of the relationship between draft and target models via KLD variance for proactive efficiency optimization.

\subsection{System-Level Optimizations}

\textbf{Large-Batch Handling.} A notable example of maintaining performance with large batch sizes is BASS~\cite{qian2024bass}. This work addresses the "ragged KV" problem, which arises from varying numbers of accepted tokens in a large batch, by introducing custom CUDA kernels (bass-pad and bass-split) implemented in DeepSpeed~\cite{rasley2020deepspeed}. While BASS improves GPU utilization for large batches, its primary limitation is the use of a fixed, batch-wide speculation length rather than a per-sequence approach.

The challenge of performance degradation in batched processing is often linked to the straggler problem, where a few slow tasks delay an entire batch. While our work identifies stragglers arising from variable speculation lengths in inference, this issue is a known bottleneck in other contexts. In their analysis of large-scale LLM training, for instance, Lin et al.~\cite{lin2025straggler} identify sequence length imbalance as a significant cause of stragglers. They attribute this to the quadratic complexity of self-attention, which causes microbatches with different sequence compositions to have vastly different processing times. Their proposed solution involves redistributing sequences across workers to balance this computational load. Our work addresses a parallel straggler problem that emerges during inference, where the imbalance is caused not by input length, but by per-sequence speculation length. Our proposed $SL_{cap}$ can therefore be seen as a novel mechanism to mitigate this specific type of inference-time straggler.

\subsection{Our Work in Context}

Existing work is thus divided: algorithmic approaches primarily use forward-looking signals like entropy for prediction, while system-level approaches manage batching inefficiencies. Our work bridges this gap by introducing a novel class of signals for dynamic adaptation. We diverge from forward-looking methods by exploring post-hoc, lagging signals—specifically using KLD to create a more robust, training-free adaptor.



\section{Methodology}
\label{sec:metho}

This study proposes a fully implemented system that extends the speculative decoding architecture of vLLM with a dynamic SL mechanism. While using the existing Draft Worker, Target Worker, and Rejection Sampler modules of vLLM, we implemented an SL Adapter module based on regional KLD features after the rejection sampling step. The SL Adapter dynamically determines the speculative length \textit{SL} for each sequence and iteration, and accordingly modifies the Look-ahead Scheduler to perform pre-mapping and reallocation of KV memory blocks. 

To further enable per-sequence decoding, our architecture integrates a variable-length kernel of \textit{FlashAttention-2}~\cite{dao2023flashattention} within the Target Worker, allowing requests with heterogeneous speculative lengths to be processed efficiently within a single batch. Importantly, the proposed architecture is not limited to conceptual design but is fully realized as an end-to-end implementation within the vLLM framework.

\begin{figure}[t]
    \centering
    \includegraphics[width=0.75\textwidth]{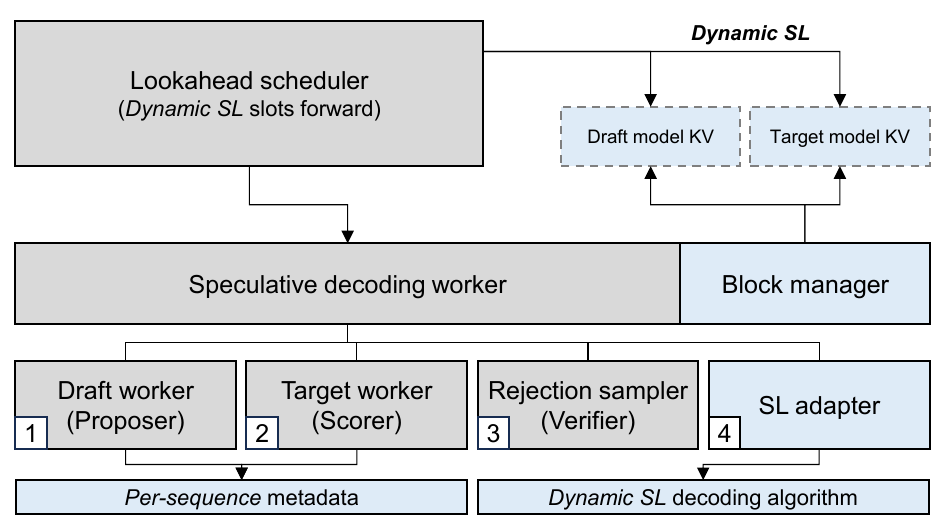}
    \vspace{-3pt}
    \caption{Workflow of the dynamic Speculative Length (SL) system. After the standard speculative decoding steps involving the Draft worker (1), Target worker (2), and Rejection sampler (3), the SL adapter (4) computes the next SL, which informs the Lookahead scheduler for the subsequent decoding round.}
    \label{fig:overview}
    \vspace{-5pt}
\end{figure}

\subsection{Dynamic SL adapter}

\subsubsection{Dynamic Calibration of Maximum Speculation Length}
To avoid a manually-tuned, fixed hyperparameter for the maximum speculation length, we introduce a lightweight pre-processing step that dynamically calibrates this value, denoted as $\hat{SL}_{max}$. This procedure avoids the significant overhead associated with training-based calibration methods. It operates by executing a small, fixed number of preliminary speculative steps at the beginning of the generation process, the performance impact of which is negligible on the overall decoding speed.

The value of $\hat{SL}_{max}$ is determined based on the statistics gathered during this brief calibration phase, according to the following formula:
\begin{equation}
    SL_{max} = SL_{A,max} \left( 1 + \frac{\mu_{KLD,pre}}{KLD_{pre,max} + \epsilon} \right)
    \label{eq:SL_max_calib}
\end{equation}
where the components are defined as follows:
\begin{itemize}
    \item $SL_{A,max}$: The maximum number of tokens accepted in any single step during the pre-processing phase. This represents an empirically observed upper bound of the model's capability.
    \item $\mu_{KLD,pre}$: The mean of all KLD values observed across all tokens during the pre-processing steps. This term reflects the average level of disagreement between the draft and target models.
    \item $KLD_{pre,max}$: The maximum single KLD value observed during the pre-processing phase. This term captures the peak disagreement.
    \item $\epsilon$: A small constant ($1 \times 10^{-6}$) added for numerical stability to prevent division by zero.
\end{itemize}
This formula allows the system to set a data-informed upper bound for the speculation length, $SL_{max}$. The ratio of mean KLD to max KLD serves as a heuristic to slightly increase the maximum length if the average disagreement is high relative to the peak disagreement, while still being anchored by the empirically observed maximum accepted length, $SL_{A,max}$.

\subsubsection{Dynamic Speculation Length Decoding Algorithm}
Based on the premise that KLD variance is a core predictive signal, we have developed a dynamic algorithm to adaptively set the $SL$, for each step. The predicted length for the next step, $\hat{SL}$, is determined by the following formula:
\begin{equation}
    \hat{SL} = (1 - SF*WVIR) \cdot (SL_{max} - SL_{min}) + SL_{min}
    \label{eq:SL_hat}
\end{equation}

\noindent where $SL_{max}$ is the effective maximum length derived from (\ref{eq:SL_max_calib}), while $SL_{min}$ is a pre-set minimum, fixed at 2. The core of this equation is the penalty term, $SF*WVIR$, which is calculated as the product of two key metrics: the Scale Factor (SF) and the Weighted Variance Intensity Ratio (WVIR).

The Scale Factor ($SF$) is designed to react to immediate model disagreement. It exponentially scales with the mean KLD of the last verified sequence, $\mu_{KLD,last}$, ensuring that recent, high-divergence steps are heavily penalized:
\begin{equation}
    SF = \exp(2 \cdot \mu_{KLD,last}) - 1
    \label{eq:sf}
\end{equation}
where $\mu_{KLD,last}$ is the mean of KLD values from the most recent step.

\begin{figure*}[!bh]
    \centering
    \includegraphics[scale=0.5]{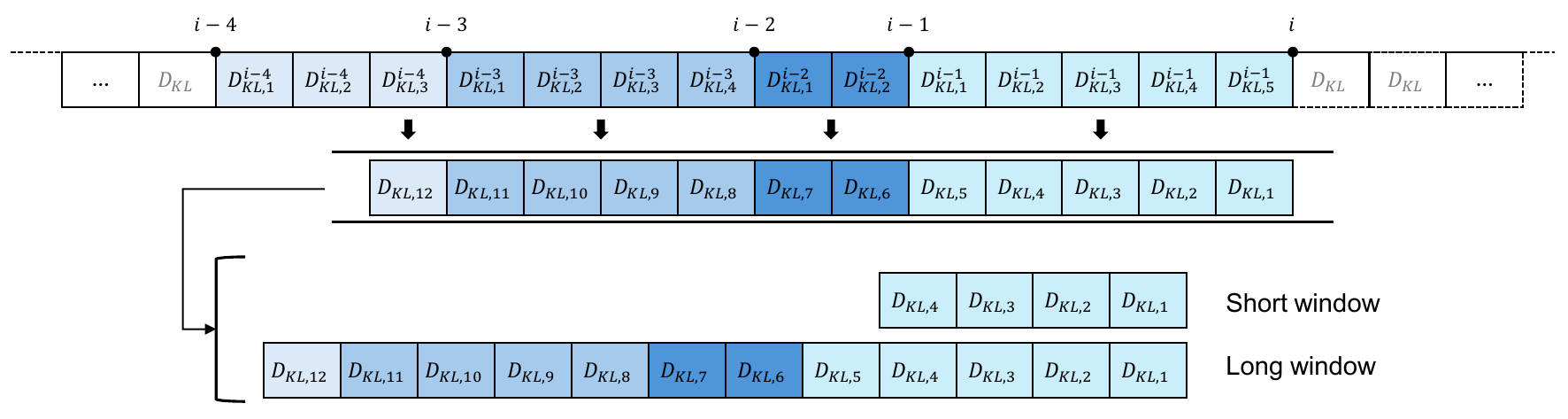}
    \caption{An illustration of the data collection process for calculating the WVIR. At each step i, the KLD values from the previous verification steps are aggregated to form distinct short-term and long-term historical windows.}
    \label{fig:kld_window}
\end{figure*}

The second component, the WVIR, measures the recent stability of the KLD signal by comparing its short-term and long-term weighted variance, calculated from historical KLD values as illustrated in Figure~\ref{fig:kld_window}. It is defined as:
\begin{equation}
    WVIR = \frac{\text{Var}_w(KLD_{short})}{\text{Var}_w(KLD_{long})}
    \label{eq:wvir}
\end{equation}
Unlike a standard variance calculation, we employ a weighted approach to assign greater importance to more recent KLD values. The weight for each step, $\alpha_i$, is determined by an exponential decay factor $\delta$ (e.g., $\delta=0.85$), where $i$ is the reverse index of the speculative step (i.e., $i=1$ is the most recent step):
\begin{equation}
    \alpha_i = \delta^{i-1}
\end{equation}
Using these weights, the weighted mean ($\mu_{w,KLD}$) and weighted variance ($\text{Var}_w$) for a given set of $N$ KLD values are calculated as follows:
\begin{equation}
    \mu_{w,KLD} = \frac{\sum_{i=1}^{N} \alpha_i \cdot KLD_i}{\sum_{i=1}^{N} \alpha_i}
\end{equation}
\begin{equation}
    \text{Var}_w(KLD) = \frac{\sum_{i=1}^{N} \alpha_i \cdot (KLD_i - \mu_{w,KLD})^2}{\sum_{i=1}^{N} \alpha_i}
    \label{eq:weighted_variance}
\end{equation}
This weighted variance is calculated separately for a short-term window ($N=10$) and a long-term window ($N=30$) to produce $\text{Var}_w(KLD_{short})$ and $\text{Var}_w(KLD_{long})$, respectively. A WVIR value greater than 1 thus indicates growing instability.

The product of SF and WVIR then forms the final penalty term, representing a synthesis of immediate (local) and historical (global) model disagreement. To ensure stability, a lower bound is enforced on this penalty. The algorithm operates under the following conditional logic:
\begin{equation}
    \hat{SL} = 
    \begin{cases}
        \begin{aligned}
            &(1 - SF \cdot WVIR) \cdot \Delta SL \\
            & \qquad + SL_{min}
        \end{aligned} 
        & \text{if } SF \cdot WVIR \leq 1 \\
        SL_{min} 
        & \text{otherwise}
    \end{cases}
    \label{eq:conditional_SL_compact}
\end{equation}
\noindent where $\Delta SL$ is $SL_{max}-SL_{min}$. This condition ensures that when the combined penalty term indicates extreme instability, the system defaults to its most conservative strategy by setting the speculation length to its minimum, $SL_{min}$. This prevents the algorithm from predicting a negative or zero length and maintains a baseline level of speculative execution.

\subsection{Per-sequence speculative decoding in vLLM}

We extend speculative decoding in vLLM~0.8.4 from batch-level static $SL$ to per-sequence dynamic $SL$. For sequence $i$ at decoding step $t$, let $SL_i^{(t)}\in\mathbb{N}$ denote the target number of speculative tokens. The design treats $SL_i^{(t)}$ as a key metadata; the scheduler allocates look-ahead work per sequence; and the verification/sampling path admits non-uniform proposal lengths in a single pass without loss of correctness. We refer to per-step, per-sequence variation of $SL_i^{(t)}$ within a batch as \emph{Ragged Q}.

At each step, the dynamic scheduler computes the next-step targets $SL_i^{(t+1)}$ from acceptance statistics; workers retrieve these targets, and the \texttt{LLMEngine} aggregates and applies them to the active queue, preserving synchronization at sequence granularity.

Scheduling uses a dedicated routine that computes lookahead slots directly from $SL_i^{(t)}$ and is applied uniformly to prefill, decode, and chunked prefill. This removes inconsistencies between feasibility checks and append operations and aligns capacity planning with intra-batch heterogeneity.

The scorer uses the \textit{FlashAttention-2} variable-length kernel, so no additional padding is required. However, verification proceeds along the $SL_{\max}^{(t)}=\max SL_i^{(t)}$ with sequence-specific validity masks, yielding accepted tokens and recoveries in one pass. The verifier uses an accepted mask and appends only the accepted tokens. Extra (bonus) tokens are handled explicitly, and a reserved padding token ID prevents invalid token identifiers from propagating when $SL_i^{(t)}$ decreases.

Dynamic $SL$ adaptation is exposed through a minimal policy interface: configuration provides an enable flag, bounds on $SL$, and activation thresholds. For each request, the policy returns $SL_i^{(t+1)}$, which is clamped per sequence under budget constraints (remaining \texttt{max\_tokens} and context capacity). Experiments are run in vLLM eager mode without CUDA Graphs (changing $SL$ would require re-capturing graphs at each step); this is a current limitation, and vLLM~v1 is expected to support piece-wise CUDA Graphs that may mitigate the overhead.

Through these modifications, we implemented the approach in the system and empirically validated per-sequence speculative decoding using the algorithm presented in this paper.

\subsection{Adaptive Speculative Length Capping}

To mitigate the straggler problem in batched per-sequence decoding, this work introduces an adaptive speculative length cap, $SL_{cap}$. This cap functions as a dynamic upper bound, applied uniformly to all predicted speculation lengths within a single batch, to prevent outlier predictions from stalling the entire process.

The determination of the cap is framed as an optimization problem. The algorithm calculates the $SL_{cap}$ that minimizes the Mean Squared Error (MSE) between the cap itself and the set of all individually predicted lengths ($\hat{SL}_i$) for the sequences in a batch.

Formally, given a set of $N$ predicted speculation lengths $\{\hat{SL}_1, \hat{SL}_2, ..., \hat{SL}_N\}$, the MSE is defined as a function of a candidate cap, $SL_{cap}$:
\begin{equation}
    \text{MSE}(SL_{cap}) = \frac{1}{N} \sum_{i=1}^{N} (SL_{cap} - \hat{SL}_i)^2
    \label{eq:sl_cap_mse}
\end{equation}

The value of $SL_{cap}$ that minimizes this loss function is found by taking a derivative of the MSE with respect to $SL_{cap}$ to zero. This derivation yields the arithmetic mean of the predicted lengths. Therefore, the final SL-Cap for the batch is calculated as:
\begin{equation}
    \frac{\partial \text{MSE}(SL_{cap})}{\partial SL_{cap}} = 0
    \label{eq:sl_cap_derivative}
\end{equation}

\begin{equation}
    SL_{cap} = \frac{1}{N} \sum_{i=1}^{N} \hat{SL}_{i}
    \label{eq:sl_cap}
\end{equation}
This method provides a principled and computationally efficient way to establish a consensus length that reduces the negative impact of high-variance predictions in a heterogeneous batch.

\section{Experiments}
\label{sec:exp}

\subsection{Experimental Setup}
\label{subsec:experimental_setup}

We used two representative draft--target pairs: LLaMA-3.1-70B-Instruct with LLaMA-3.2-1B-Instruct, 
and Gemma-27B with Gemma-2B. We evaluated robustness and generalizability on eight diverse datasets: 
CNN/DM~\cite{hermann2015cnndm}, XSum~\cite{narayan2018xsum}, GSM8K~\cite{cobbe2021gsm8k}, HotpotQA~\cite{yang2018hotpotqa}, NQ~\cite{kwiatkowski2019nq}, HumanEval~\cite{chen2021humaneval}, 
ShareGPT~\cite{sharegpt}, and WMT14~\cite{bojar2014wmt14}. The primary metric is end-to-end request latency, measured as the average completion time of 128 prompts executed under varying batch sizes. 
Experiments were run on a single server with 8× NVIDIA A100 80GB GPUs.


\subsection{Evaluation of Predictive Signals}
\label{subsec:dyn-sl-eval}
This evaluation begins with a token-wise analysis of the correlation between potential signals and the final acceptance rate. We use the CNN/DM dataset, chosen as a task of moderate task complexity, and compute Pearson correlation coefficients between acceptance and three signals: 
the forward-looking entropy of the draft distribution, the lagging mean KLD over the previous 10 steps, and the corresponding WVIR value.

Table~\ref{tab:correlation_results} presents the Pearson correlation coefficients between these signals and the token acceptance rate at sampling temperatures of 0.0 and 1.0. The results show that all evaluated signals exhibit a weak correlation with token-level acceptance, and this relationship weakens further when moving from near-greedy decoding to stochastic sampling.

\begin{table}[t]
\centering
\tiny 
\setlength{\tabcolsep}{4pt} 
\renewcommand{\arraystretch}{1.25} 
\caption{Pearson correlation ($r$) between signals and token acceptance probability at different sampling temperatures on the CNN/DM dataset.}
\resizebox{0.9\textwidth}{!}{%
\begin{tabular}{|p{3.2cm}|p{3.2cm}|p{3.2cm}|}
    \hline
    \textbf{Signal / Metric} & \textbf{Correlation (Temp 0.0)} & \textbf{Correlation (Temp 1.0)} \\
    \hline
    Entropy (draft) & -0.339 & -0.235 \\
    \hline
    Mean KLD & -0.164 & -0.069 \\
    \hline
    WVIR & 0.128 & -0.031 \\
    \hline
\end{tabular}
}
\label{tab:correlation_results}
\end{table}

At temperature 0.0, the forward-looking signal entropy shows a modest negative correlation ($r = -0.339$), while at temperature 1.0, sampling randomness further weakens the predictive power of all signals. 

For lagging signals such as KLD and WVIR, token-level correlations approach zero. This weak correlation arises because the optimal speculation length is itself highly volatile, 
fluctuating dramatically between iterations, which makes it difficult for any single feature to capture. For post-hoc signals, the challenge is compounded by their inherent one-step delay, as they rely on information from past steps. Thus, while ill-suited for precise token-level prediction, these signals remain valuable as macroscopic diagnostics: in particular, the variance of KLD reflects regional stability and provides guidance for speculation length adaptation.

\subsection{End-to-End Performance and Robustness}

\label{subsubsec:perfom-latency}

Given the weak token-level correlation of the available signals, we next evaluate the end-to-end performance of our proposed algorithm. The focus here is not on direct prediction, but on whether diagnostic signals can drive robust and stable performance across diverse workloads.

Obtaining the static-opt baseline is computationally costly: profiling five SL values (2, 4, 6, 8, 10) per dataset required 9825 seconds (~2.7 hours) on average, totaling nearly 22 hours across eight datasets. 

Table~\ref{tab:latency_subtables} reports results for the LLaMA-3.1-70B-Instruct (target) with LLaMA-3.2-1B-Instruct (draft). At temperature 0.0, all accelerated methods achieved substantial speedups over the incremental baseline (38.41s). Our algorithm reached 13.97s, comparable to AdaEDL (13.83s) and close to static-opt (13.44s), but without profiling overhead. 

At temperature 1.0, performance remained competitive (19.19s vs. 18.02s static-opt and 17.64s AdaEDL), though the added sampling randomness slightly widened the gap. We hypothesize that this modest performance lag is attributable to the nature of our signal; the increased randomness inherent in sampling may introduce token-level noise that is unfavorable for an algorithm that relies on lagging, regional signals to diagnose stability. Overall, the proposed method offers latency on par with leading baselines, while avoiding costly per-dataset profiling. 

\begin{table}[t]
\centering
\tiny 
\setlength{\tabcolsep}{4pt}
\renewcommand{\arraystretch}{1.3}
\caption{Latency results and speed-ups compared to the autoregressive decoding 
for LLaMA3.1-70B-Instruct (target) with LLaMA3.2-1B-Instruct (draft). 
Values are in seconds.}
\label{tab:latency_subtables}

\begin{minipage}[t]{0.48\textwidth}
\centering
\textbf{(a) Temperature 0.0}\\[2pt]
\begin{tabular}{|p{2.7cm}|p{2.0cm}|p{1.0cm}|}
\hline
\textbf{Method} & \textbf{Mean Latency (s)} & \textbf{Speedup} \\
\hline
Autoregressive      & 38.41 & 1.00× \\
\hline
Static-opt          & 13.44 & 2.86× \\
\hline
Proposed Dynamic SL & 13.97 & 2.75× \\
\hline
AdaEDL (base=7)     & 13.83 & 2.78× \\
\hline
\end{tabular}
\end{minipage}
\hfill
\begin{minipage}[t]{0.48\textwidth}
\centering
\textbf{(b) Temperature 1.0}\\[2pt]
\begin{tabular}{|p{2.7cm}|p{2.0cm}|p{1.0cm}|}
\hline
\textbf{Method} & \textbf{Mean Latency (s)} & \textbf{Speedup} \\
\hline
Autoregressive      & 38.47 & 1.00× \\
\hline
Static-opt          & 18.02 & 2.13× \\
\hline
Proposed Dynamic SL & 19.19 & 2.00× \\
\hline
AdaEDL (base=7)     & 17.64 & 2.17× \\
\hline
\end{tabular}
\end{minipage}

\end{table}

\begin{figure*}[t]
    \centering
    \includegraphics[width=1.0\textwidth]{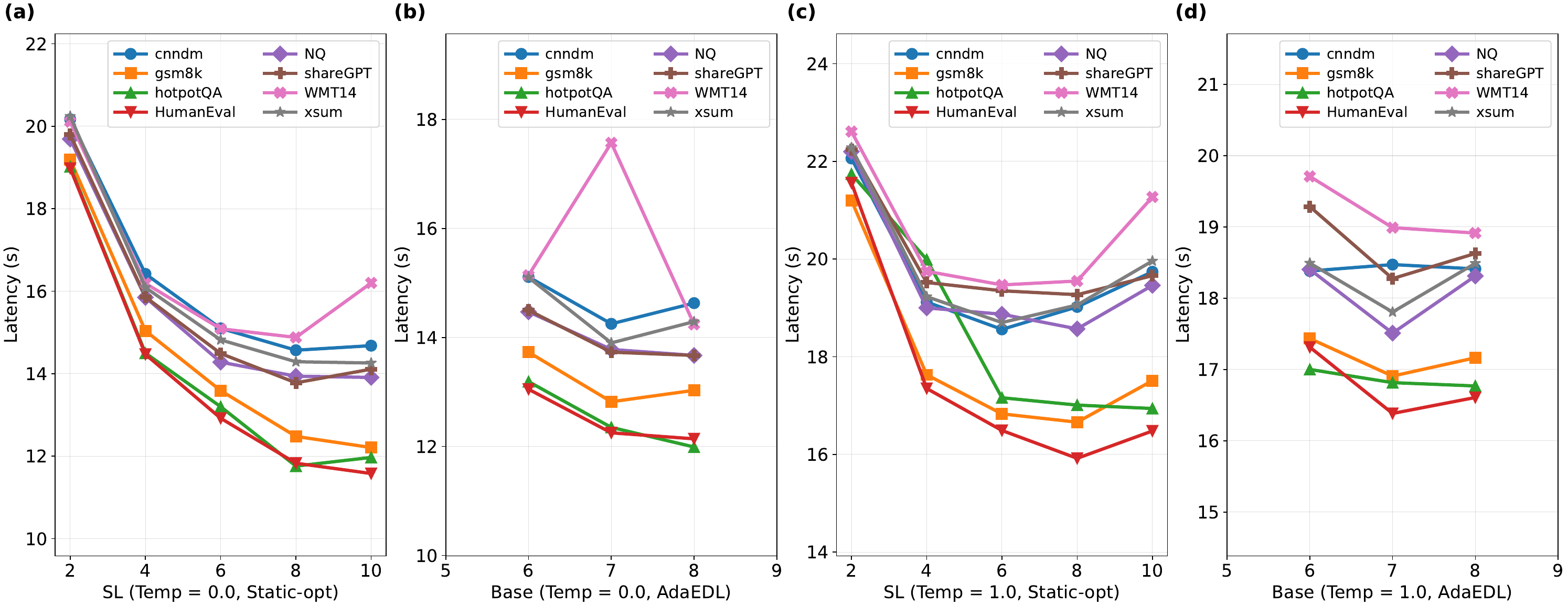}
    \caption{Performance sensitivity to speculation length hyperparameters. The figure compares the latency of static methods across different speculation lengths with dynamic methods (AdaEDL) across a range of maximum speculation lengths, evaluated at temperatures 0.0 and 1.0.}
    \label{fig:sens_latency}
\end{figure*}

The analysis highlights the hyperparameter sensitivity of baseline methods. Static-SL exhibits strong hyperparameter sensitivity, with a U-shaped latency curve (Fig.~\ref{fig:sens_latency}) 
and sharp degradation outside its optimum. AdaEDL shows weaker but still noticeable sensitivity to its base setting, indicating that achieving peak performance still depends on careful tuning.

In contrast, our proposed method shows notable robustness by leveraging post-hoc KLD-based diagnostics. As shown in Fig.~\ref{fig:across_dataset}, it consistently matches the per-dataset static-opt baseline and AdaEDL across diverse datasets, while avoiding costly per-dataset profiling.

\subsection{Analysis in a Low-Acceptance-Rate Regime}

To evaluate the robustness of the dynamic algorithms under more challenging conditions, this section analyzes their performance in a low-acceptance-rate regime. This scenario was created using a Gemma-27B target model and a Gemma-2B draft model, a pair with significant divergence. The high level of disagreement resulted in an extremely conservative optimal static speculation length ($k_{opt}=2$) for most datasets, indicating a difficult environment for speculative decoding.

\begin{figure}[t]
    \centering
    \begin{minipage}[t]{0.48\textwidth}
        \centering
        \includegraphics[width=\linewidth]{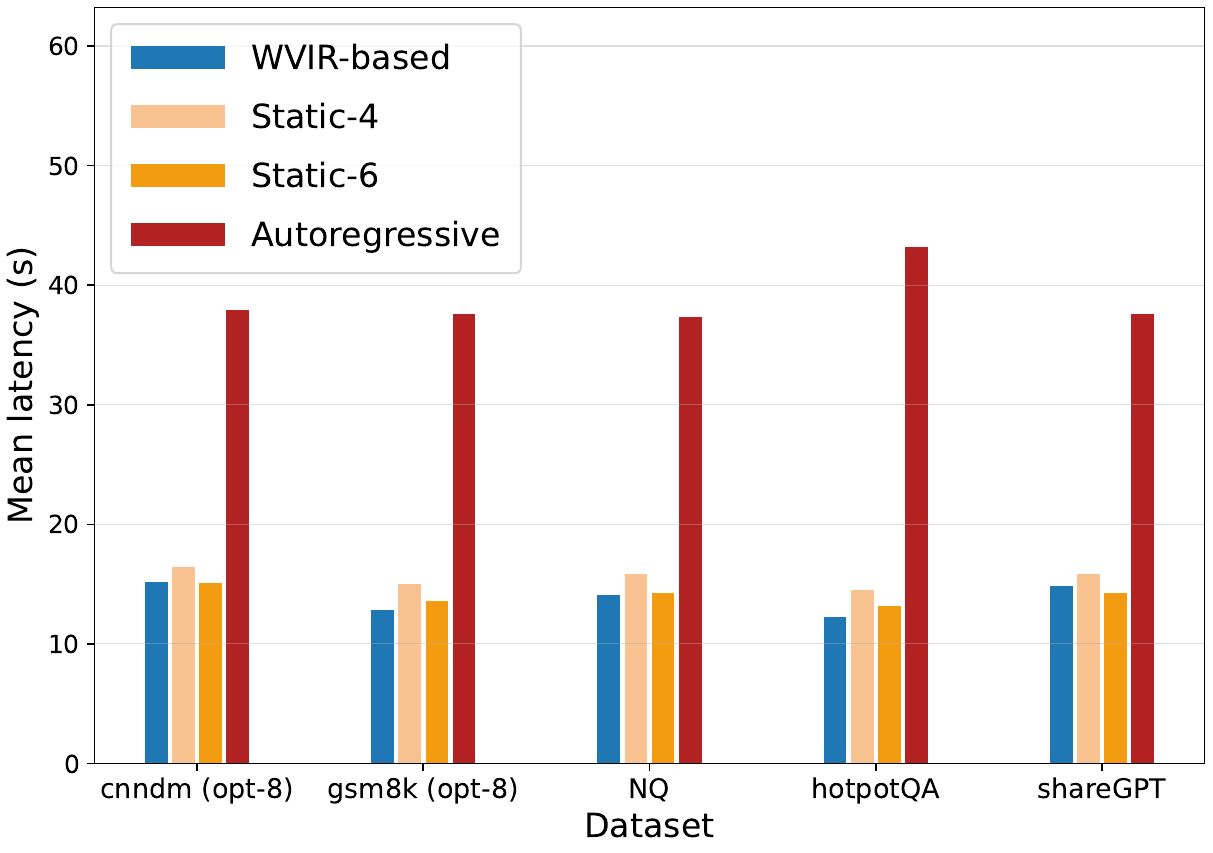}
        \vspace{-3pt}
        \caption{Mean latency of the WVIR-based algorithm compared to AdaEDL 
        and the per-dataset Static-opt baseline across datasets at temperature 0.0.}
        \label{fig:across_dataset}
    \end{minipage}
    \hfill
    \begin{minipage}[t]{0.48\textwidth}
        \centering
        \includegraphics[width=\linewidth]{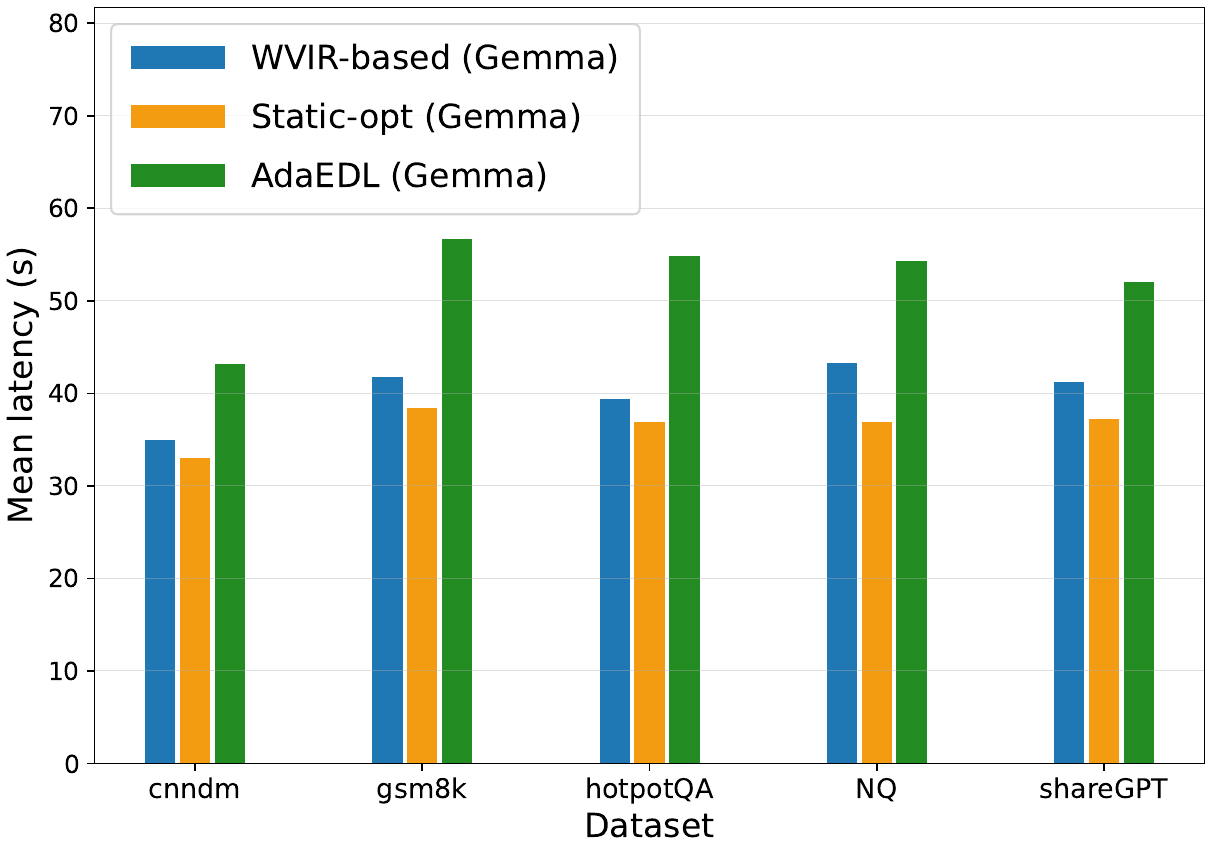}
        \vspace{-3pt}
        \caption{Mean latency of dynamic methods and the Static-opt baseline 
        in a low-acceptance-rate regime using Gemma models.}
        \label{fig:extreme_case_latency}
    \end{minipage}
    \vspace{-5pt}
\end{figure}

Figure~\ref{fig:extreme_case_latency} presents the mean latency of each dynamic method under these adverse conditions. The results reveal a notable divergence in performance. While the proposed WVIR-based algorithm maintained a latency close to the static-opt baseline, AdaEDL's performance degraded substantially.

The relative performance degradation is detailed in Table~\ref{tab:extreme_case_latency}, which presents the latency for the Gemma model pair, normalized by the latency achieved with the LLaMa3.1-70B/LLaMa3.2-1B model pair. The results reveal a notable divergence in performance under these adverse conditions. While the proposed WVIR-based algorithm maintained a latency close to the Static-opt baseline (e.g., 234\% the static-opt latency on CNNDM), AdaEDL's performance degraded substantially.

\begin{table}[t]
\centering
\tiny 
\setlength{\tabcolsep}{4pt} 
\renewcommand{\arraystretch}{1.3} 
\caption{Percentile increment in mean latency of dynamic methods with Gemma 27B/2B, compared to the LLaMA-70B/1B.}
\resizebox{0.9\textwidth}{!}{%
\begin{tabular}{|p{2.3cm}|p{1.8cm}|p{1.8cm}|p{1.8cm}|}
    \hline
    \textbf{Dataset} & \textbf{Static-opt} & \textbf{AdaEDL} & \textbf{WVIR-based} \\
    \hline
    CNNDM & 178\% & 234\% & 180\% \\
    \hline
    GSM8K & 231\% & 335\% & 234\% \\
    \hline
    NQ & 199\% & 310\% & 229\% \\
    \hline
    ShareGPT & 191\% & 285\% & 208\% \\
    \hline
    WMT14 & 194\% & 284\% & 198\% \\
    \hline
\end{tabular}
}
\label{tab:extreme_case_latency}
\end{table}

A hypothesis for this performance divergence lies in the nature of the signals used by each algorithm, especially under conditions of severe model disagreement. Methods relying primarily on forward-looking signals, such as the draft model's entropy, face a potential information imbalance; their decisions may be biased by the draft model's own confidence, which is frequently incorrect in this regime.

While a method like AdaEDL attempts to account for this by incorporating the historical acceptance rate, this metric is itself a consequential outcome of the process rather than a foundational measure of disagreement. In contrast, a lagging signal like KLD, by definition, incorporates information from both the draft and target models. This access to a more complete picture of the model dynamics may offer a more robust, albeit delayed, diagnostic. This suggests a potential advantage for KLD-based signals in high-divergence, low-acceptance-rate scenarios, where signals based predominantly on the draft model's state become less reliable.

\begin{figure}[t]
    \centering
    \includegraphics[width=0.63\textwidth]{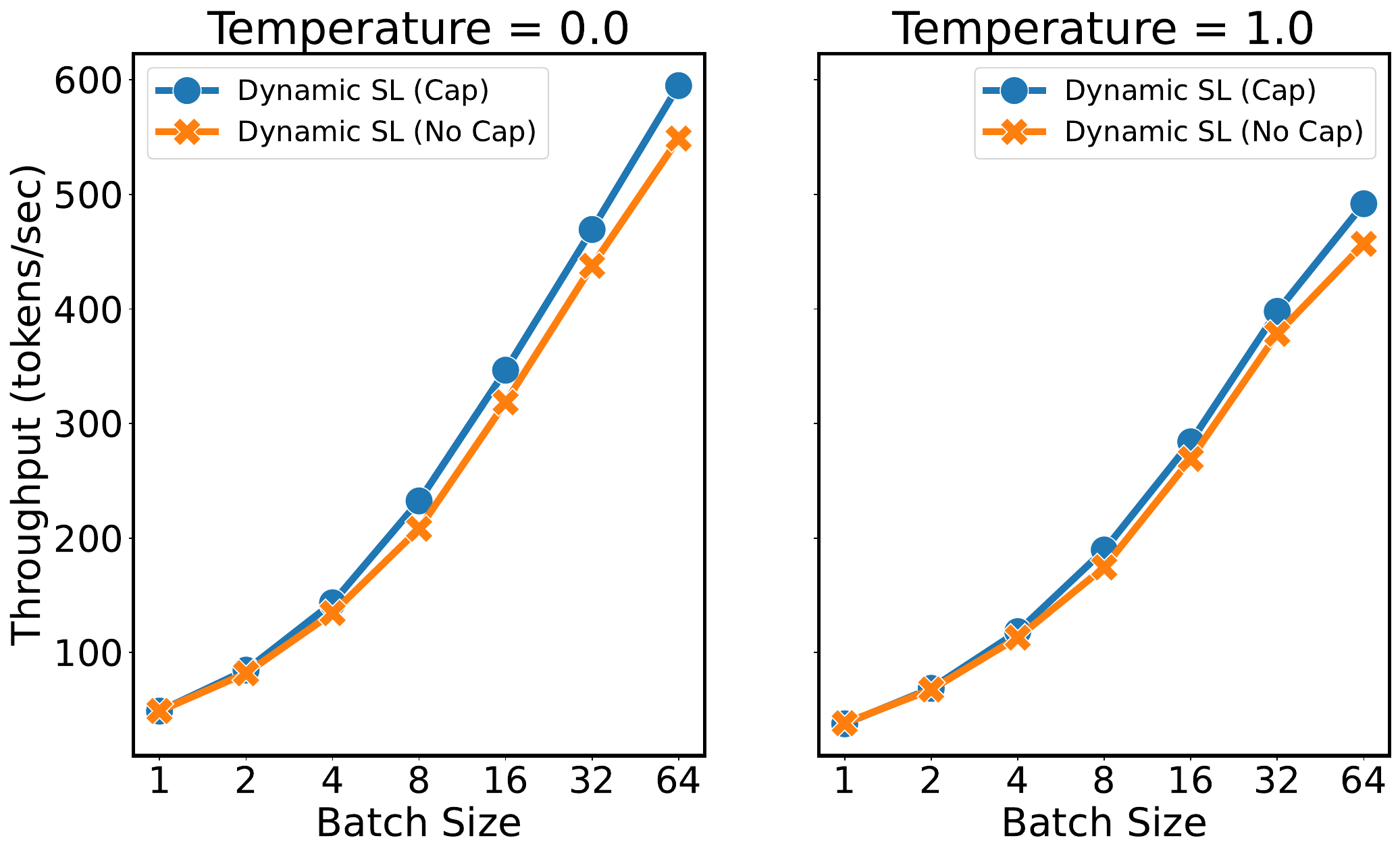}
    \vspace{-3pt}
    \caption{Throughput scalability of speculative decoding strategies across different batch sizes, 
    evaluated under greedy decoding (Temperature=0.0, left) and stochastic sampling (Temperature=1.0, right).}
    \label{fig:SL_cap_test}
    \vspace{-5pt}
\end{figure}

\subsection{Efficacy of Per-Sequence Adaptation and $SL_{cap}$}
This section evaluates the scalability of the per-sequence approach and the efficacy of the $SL_{cap}$ in mitigating its limitations. To conduct this experiment, we used the CNN/DM dataset and tested batch sizes from 1 to 64, measuring the corresponding throughput in tokens per second. In an ideal scalable system, throughput should increase linearly with batch size. However, the results in Figure~\ref{fig:SL_cap_test} demonstrate that the naive per-sequence strategy (\texttt{Dynamic SL (No Cap)}) exhibits suboptimal scalability as the batch size increases. This scalability gap is consistent across temperatures. Compared to its performance at a batch size of 1, the \texttt{No Cap} strategy's throughput scales by only 11.21x at temperature=0.0 and 11.92x at temperature=1.0. This degradation is caused by the "straggler problem" as mentioned in Section~\ref{sec:Introduction}: a few sequences with aggressively long predicted speculation lengths stall the entire batch. This finding highlights that while a naive per-sequence approach is theoretically appealing, it suffers from practical scalability issues in large-batch environments.

The effectiveness of the $SL_{cap}$ is clearly demonstrated by comparing the capped and uncapped strategies in Figure~\ref{fig:SL_cap_test}. The application of the $SL_{cap}$ substantially mitigates the performance degradation. With the cap applied, the throughput at a batch size of 64 improves by 12.16x (temperature=0.0) and 13.01x (temperature=1.0) relative to a batch size of 1. This shows a marked improvement in scalability over the uncapped approach, allowing the system's performance to scale more effectively with larger batch sizes. The cap successfully curtails outlier predictions, preventing the straggler problem from dominating batch processing time. This result highlights that a regulation mechanism like the $SL_{cap}$ can be a main feature for mitigating performance degradation in dynamic, per-sequence speculative decoding for real-world, high-throughput serving systems.

\section{Conclusion}
\label{sec:con}

In this work, we explored the critical challenge of using a static SL in speculative decoding for large-batch LLM serving. We introduced \sysname, a novel two-level dynamic adaptation system designed to investigate the potential of a new class of predictive signals. Our approach is training-free and combines two main contributions: a predictive algorithm that uses the variance of post-hoc KLD as a signal for regional stability, and an adaptive SL cap that resolves the straggler problem inherent in per-sequence decoding.
%

Our experimental evaluation validates this approach. While a correlation analysis confirmed that direct, token-level prediction from signals like KLD is challenging, the end-to-end results demonstrate the value of our approach. Across a diverse set of eight datasets, the WVIR-based dynamic SL algorithm achieved average latency competitive with both per-dataset static-optimal baselines and other dynamic methods like AdaEDL. The significance of our approach was most evident in its robustness; it showed greater performance stability across different datasets and less sensitivity to hyperparameter choices. Furthermore, in a high-divergence, low-acceptance-rate regime where entropy-based signals faltered, our KLD-based method maintained stable performance. Finally, we demonstrated that our adaptive SL cap successfully mitigates the straggler problem in large-batch serving, substantially improving throughput scalability.

The primary direction for future work is to refine the predictive model based on KLD variance. While our study validates its potential as a signal, further feature engineering and combination with other heuristics could lead to significant performance gains. Another crucial avenue for improvement involves addressing a current implementation limitation: our system operates in vLLM's eager mode, which precludes the performance benefits of CUDA Graphs. Integrating our dynamic logic with emerging features like piece-wise CUDA Graphs will be a key next step to reduce overhead.  

Beyond speculative length, per-sequence decoding could also be extended to incorporate sequence-specific hyperparameters (e.g., temperature, repetition penalty), which are currently fixed across all requests in batch inference regardless of their contextual differences. Adapting such parameters on a per-sequence basis provides another axis of optimization, enabling serving systems to better handle heterogeneous workloads in terms of both efficiency and output quality.  

Ultimately, \sysname serves as a proof-of-concept, demonstrating that post-hoc signals, in conjunction with per-sequence adaptation and context-aware hyperparameter control, offer a promising and underexplored path for creating the next generation of adaptive speculative decoding systems.

\bibliographystyle{unsrt}      
\bibliography{references}

\end{document}